\documentclass[twocolumn]{aastex631}

\newcommand{\diff}{\delta}
\newcommand{\diffpstar}{\delta_{p*}}
\newcommand{\tse}{{t}_{se}}

\newcommand{\tsehat}{\hat{t}_{se}}

\begin{document}

\title{Observability of flashes from ejecta crashes in aspherical supernovae, with application to SN 2008D} 

\author[0009-0005-0313-3129]{Benjamin Scully}
\affiliation{Department of Astronomy and Astrophysics, University of Toronto,\\
50 St. George Street, Toronto, ON M5S 3H4, Canada} \email{ben13@student.ubc.ca}
\affiliation{Department of Physics and Astronomy, University of British Columbia\\
325 - 6224 Agriculture Road, Vancouver, BC V6T 1Z1, Canada}
\author[0000-0001-9732-2281]{Christopher D. Matzner}
\affiliation{Department of Astronomy and Astrophysics, University of Toronto,\\
50 St. George Street, Toronto, ON M5S 3H4, Canada} \email{matzner@astro.utoronto.ca}

\author[0000-0002-5230-5514]{Almog Yalinewich}
\affiliation{vfunction, R\&D Center, Hapelech 7, Tel Aviv, Israel, 6816727}

\begin{abstract}
A new class of transient, which has been hypothesized to accompany the explosion of an aspherical compact supernova, would arise when streams of ejecta collide outside the star.  However, conditions that favour the prompt release of radiation from the collision, such as a diffuse stellar envelope, disfavour the creation of non-radial ejecta in the first place.  To determine whether the collision can both occur and be visible, we simulate aspherical explosions using the HUJI-RICH moving-mesh hydrodynamics code and analyze them in terms of diffusion measures defined for individual fluid elements.  While our simulations are highly idealized, they connect to realistic explosions via a single dimensionless parameter.   Defining two measures of the importance of diffusivity (two versions of the inverse P\'eclet number), we find that one varies in a way that indicates colliding ejecta can release a photon flash, while the other does not. Examining the x-ray transient XT\,080109 associated with supernova SN\,2008D, we find that its fluence and duration are consistent with the properties of an ejecta collision in the aspherical model that is most likely to emit a flash. Our results give tentative evidence for the possibility of collision-induced flashes for a narrow and radius-dependent range of asphericity, and motivate future radiation hydrodynamics simulations.

\end{abstract}


\section{Introduction} \label{sec:intro}

Stripped-envelope supernovae (SESNe) are expected to exhibit unusual behavior in their shock breakout and early light curve, thanks to a particular combination of qualities:   (1) higher energy-to-ejected-mass ratios, especially among the broad-lined Ic subclass \citep{hamuy2003observed}; (2) more compact stellar envelopes in which radiation diffusion is suppressed, allowing for a greater degree of shock acceleration in the star's sub-surface density gradient; and (3) larger deviations from spherical symmetry, as determined from polarization \citep{wang2001bipolar,len05} and from nebular-phase spectra \citep{2022ApJ...928..151F}.   The first two properties give especially energetic SESNe the unique capacity to create relativistic ejecta, even in spherical symmetry \citep{mm99,2001ApJ...551..946T}, but also make their accelerating shock especially sensitive to non-spherical perturbations \citep{M13,Irw21}.  Given that the same events are typically aspherical, this sensitivity implies that SESNe should show the hallmarks of non-spherical behavior in their fastest ejecta and in their earliest light \citep{cou11,Afs18}.  

Of the likely hallmarks, a particularly dramatic possibility involves collisions between streams of ejecta outside the star.     Such collisions occur for explosions that are sufficiently compact, energetic, and aspherical, as we explain below -- and they might produce flares analogous to those expected from ejecta-companion interactions in Type Ia supernovae \citep{2010ApJ...708.1025K}, as discussed by \citet{M13}, \citet{S14}, \citet{Afs18}, \citet{Irw21} and \citet{DuP22}.  If so, the radiation is most likely to emerge as a $\sim10^{2-3}$\,s flare of total energy $\sim 10^{46-47}$\,erg, most likely in the soft x-ray band, although reprocessing to lower-energy photons is also possible. 

On the other hand, ejecta-ejecta collisions might not produce prompt observable flares at all.  The fundamental problem is that the qualities of an explosion that enable ejecta collisions to exist (by suppressing radiation diffusion as non-radial pressure gradients develop) can also stifle the flare (by suppressing the escape of radiation as the ejecta collide).  {\em  Can one have a collision and see it too?}   This is the question we seek to answer. 

Let us suppose that some global parameter $\diffpstar$ captures how strongly radiation diffusion affects the creation of non-radial ejecta. (We shall define $\diffpstar$ more precisely in \S\,\ref{S:scalings}.) When $\diffpstar$ is sufficiently large, non-radial ejecta are never produced, and no collision exists.  Conversely, when $\diffpstar$ is sufficiently small, a non-spherical explosion launches streams of ejecta onto colliding trajectories, but no flare can emerge from the collision as it happens. 
(In this case the collision imprints itself in the ejecta density and ejecta, in ways that might affect the later light curve or circumstellar interactions.)   Our question, therefore, is whether there exists any gap between these regimes: some intermediate range of $\diffpstar$ for which the collision both occurs, and also promptly releases its energy as light. 
 
As a useful first step toward answering this question, we create a set of adiabatic simulations in order to  evaluate the importance of photon diffusion in fluid parcels as they accelerate and collide.  Our approach provides an estimate of  which values of the explosion parameters, if any, are most likely to produce collision flares.  Our simulations are very similar to those of \cite{Afs18}, except that we employ the Lagrangian code HUJI-RICH \citep{rich15}.   This provides a means to achieve variable resolution, permitting simulations on personal computers, and to facilitate the tracking of diagnostics along fluid trajectories, a key feature of our analysis. 

We begin by briefly reviewing the dynamics of spherical and aspherical explosions, and examining the importance of radiation diffusion on the equatorial collision of ejecta. 

\section{Scalings and Expectations} \label{S:scalings}

We start with the spherical case.  Consider an idealized spherical explosion of stellar radius $R$, ejected mass $M$ (with radial mass and density distributions $m(r)$ and $\rho_0(r)$, respectively), and energy $E$ -- parameters that define a density scale $\rho_*=M/R^3$, a  velocity scale $v_*=(E/M)^{1/2}$, and a time scale $t_*=R/v_*$.   The character of this explosion is defined by  dimensionless ratios defining the strength of gravity ($\Pi_{\rm grav} = GM_\star^2/ER$, where the gravitating mass $M_\star$ might exceed $M$),  of gas pressure ($\Pi_{\rm gas} = h^3c^3M^4/E^3R^3\mu^4$, where $\mu$ is the mean molecular weight), of relativity ($\Pi_{\rm rel}=E/Mc^2$), and  -- especially relevant to this study -- of radiation diffusion, 
\begin{equation} 
\Pi_{\rm diff} = \frac{c R^2}{\kappa\sqrt{EM}},
\end{equation} 
where $\kappa$ is the specific opacity relevant to the explosion (usually, from electron scattering). For our simulations we adopt adiabatic, non-gravitating flow, corresponding to the strong, weakly diffusive limit $\Pi_{\rm grav}\ll1$ and $\Pi_{\rm diff}\ll1$.  (Strong explosions are weakly diffusive because diffusion is already slow compared to diffusion in the progenitor star.)   We also neglect gas pressure ($\Pi_{\rm gas}\ll1$) which is appropriate unless the progenitor is very compact ($R\lesssim R_\odot$), and even then, radiation pressure tends to dominate the outer regions that produce colliding flows.  

In spherical symmetry, the progress of a strong but non-relativistic  explosion is captured by the \cite{mm99} shock velocity approximation: 
\begin{equation} 
v_s(r) = A v_* \tilde{m}(r)^{-1/2} \left[\frac{\tilde{m}(r)}{\tilde{r}^3 \tilde{\rho}_0(r)}\right]^{\beta}
\end{equation} 
in which $A \simeq 0.794$ and $\beta \simeq 0.19$, but see \cite{2001ApJ...551..946T} and \cite{2013ApJ...773...79R} for refinements. Tildes indicate normalized quantities: $\tilde \rho_0=\rho_0/\rho_*$, $\tilde r = r/R$, $\tilde m = m/M$, etc.   In this explosion, the shock emerges from the stellar surface after a time $\tse =\int_0^R v_s(r)^{-1} dr$.   

In an {\em aspherical}  explosion, shock eruption is not simultaneous: the time at which the shock emerges at every point on the stellar surface, $\tse(\theta)$ (labeled by polar angle $\theta$, assuming axisymmetry), is not constant.  Its inverse gradient along the surface defines a local pattern velocity ${\mathbf v}_p(\theta)$.  This lack of simultaneity, coupled with the tendency for a shock to accelerate in the strong density gradient near the stellar surface, causes the shock front to curve toward the surface -- that is, the shock normal turns away from the radial direction, becoming horizontal (parallel to $\mathbf{v}_p$) at a  characteristic depth $\ell_p(\theta)$ for which the radial shock speed matches the local pattern speed (see \citealt{M13} figure 1).  

It is useful to define, for reference, a spherical analog to the  aspherical explosion under consideration: one with the same $E$,  $R$, $M$, and $\rho_0(r)$.  We refer to the shock velocity and breakout time of this spherical analog as $\hat v_s(r)$ and $\tsehat$, respectively. 
Estimating the radial velocity within our aspherical explosion as $[\tsehat/\tse(\theta)] \hat{v}_s(r)$ gives
\begin{equation} \label{eq:ell_p-condition}
  \hat{v}_s(r=R-\ell_p(\theta)) = \frac{\tse(\theta)}{\tsehat} v_p(\theta),
 \end{equation} 
 from which $\ell_p(\theta)$ can be determined.    Matter excavated from depths greater than $\ell_p(\theta)$ flies away in a mostly radial direction, while shallower material will be cast increasingly sideways.  
 
However, the diffusion of radiation can prevent non-radial motions because of the importance of photon pressure in the post-shock flow.  Considering the photon diffusivity $\nu = c/3\kappa\rho$, the effect of diffusion on the non-radial flow is determined by the dimensionless ratio (inverse P\'eclet number) 
\begin{equation} \diff_p(\theta) = \frac{c}{3\kappa \rho_p(\theta) \ell_p(\theta) v_p(\theta) }, \end{equation} 
which we call the `oblique diffusion parameter'. 
Wherever $\diff_p(\theta) \ll 1$, non-radial ejection is possible, and conversely, wherever $\delta_p(\theta)\gg1$, diffusion erases the necessary pressure gradients.  In the latter case, the shock suffers diffusive breakout in a layer below $\ell_p(\theta)$, where motion is still mostly radial.    

 To characterize the influence of diffusion on non-radial flows across an entire explosion, we note that  $\tilde\ell_p(\theta)$ and $\tilde\rho_p(\theta)$ are functions of $\tilde v_p(\theta)$ (given the progenitor structure, $\tilde\rho_0(\tilde r)$) -- and so $\diff_p(\theta)$ is as well.   Focusing on bipolar explosions, we define the pole-to-equator  time lag for shock eruption to be $\tse(\pi/2)-\tse(0) = 2\epsilon \tsehat$ (thereby defining $\epsilon$ in a manner similar to \citealt{M13}).  The time-averaged pattern speed is  
 \begin{equation} \bar v_p = \frac{\pi}{4\epsilon} \frac{t_*}{\tsehat} v_*,\end{equation} and we define the `global oblique diffusion parameter', $\diffpstar$, to be $\diff_p$ evaluated with $v_p \rightarrow \bar v_p$ and $\tse\rightarrow\tsehat$.   In general, $\diff_{p*}$ equals $\Pi_{\rm diff}$ times some function of $\epsilon$ that depends on the stellar structure, as we demonstrate below for polytropic envelopes. 
 
The point of this is that $\diffpstar$ provides a way to assess the one aspect of aspherical explosions that we aim to study, even for explosions that differ in other ways.  This also justifies the shortcut we employ, in which we  we analyze a single adiabatic simulation of an  aspherical explosion.  Adopting a non-zero opacity in post-processing allows us to assess where diffusion would affect the flow for a given value of $\diffpstar$, determining in the process whether diffusion is more important within the collision region than during ejection. 

Why should $\diffpstar$ describe the diffusivity of the collision, as well as the eruption process?   This relies on the point, elaborated first by \citet{M13}, that the influence of diffusion is approximately constant for each mass element as it flies away from the ejection site, at least until it has traveled a distance comparable to the stellar radius.  Because the relevant collisions occur at radii a couple times $R$, we expect diffusion to be marginally more relevant to the collision than it is during ejection.   Whether this logic holds within a real explosion, and whether it implies that collisions can release a burst of photons, is the subject of our study. 

\subsection{Polytropic stellar envelopes}

In the outer reaches of a polytropic envelope of index $n$, where the enclosed mass is nearly constant, the hydrostatic structure approaches the definite form 
\begin{equation}  \label{eq:rho0_Polytrope}
\tilde{\rho}_0 = \tilde{\rho}_h \left(\frac{1}{\tilde{r}}-1\right)^n,
\end{equation}
 where $\rho_h$ is the density extrapolated to a reference radius of $\tilde r=1/2$.  This limit provides an accurate approximation, holding to within roughly $3\%$ for $\tilde{r}>0.6$ when $n=3$.   
 (Its limiting power law form $\tilde\rho_0 \rightarrow \tilde\rho_h (1-\tilde r)^n$, in contrast, is only accurate in the outer few percent of $R$.)   Within this outer region of validity, $\tilde m\simeq 1$ and thus 
 \begin{equation} 
\hat v_s \simeq A v_\star \tilde\rho_h^{-\beta} \tilde{r}^{(n-3)\beta} (1-\tilde r)^{-n\beta}.
 \end{equation} 

At this point we focus on the case $n=3$, for which $\beta = 0.185747$ \citep{2013ApJ...773...79R}, both because this simplifies the evaluation and because it describes our simulations.   Condition (\ref{eq:ell_p-condition}) implies 
\begin{equation}
\tilde \ell_p(\theta) = \left[ \frac{A \tsehat}{\tse(\theta)\tilde v_p(\theta)}\right]^{1/3\beta} \tilde\rho_h^{-1/3},  
\end{equation} 
and our definition of the global diffusion parameter gives 
\begin{eqnarray}\label{eq:diffpstar}
\diffpstar &=& 
     \frac13 \Pi_{\rm diff}  \tilde\rho_h^{1/3} \left(\frac{\bar v_p}{v_*}\right)^{4/3\beta -1} (1-\tilde\ell_{p*})^3
        \nonumber\\
     &=& \frac{ \Pi_{\rm diff} }{(1.521 \epsilon)^{6.178} } \tilde\rho_h^{1/3} (1-\tilde\ell_{p*})^3,
\end{eqnarray}
so long as $\tilde\ell_{p*}\lesssim 0.4$, where the quantity
\begin{equation} 
\tilde\ell_{p*} = \left( \frac{4\epsilon A \tsehat}{\pi t_*}\right)^{1/3\beta} \tilde\rho_h^{-1/3} \\ 
\end{equation} 
estimates the relative depth at which the shock becomes strongly non-radial.  

To summarize: our global parameter $\Pi_{\rm diff}$ indicates the strength of radiation diffusion during the explosion overall, while the adjusted $\diffpstar$ evaluates diffusion specifically for elements caught up within the oblique and non-radial flow. Note that the stellar radius has an appreciable influence on the oblique diffusion parameter ($\diffpstar\propto\Pi_{\rm diff}\propto R^2$), but non-sphericity has a much stronger influence ($\diffpstar \propto \epsilon^{-6.178}$).   Although these specific formulae and scalings assume an $n=3$ polytrope, we anticipate that similar results will hold for more realistic stellar structures.

\section{Simulation} \label{sec:sim}

We closely follow \cite{Afs18} in defining our progenitor and its explosion.  We consider the case of a strong, radiation dominated, non-relativistic explosion ($\Pi_{\rm grav}, \Pi_{\rm gas}, \Pi_{\rm rel}$, all $\lll1$) and conduct our numerical simulations without radiation diffusion (corresponding to $\Pi_{\rm diff}\lll1$) although we account for finite diffusion in our analysis.  Our simulations therefore account only for adiabatic fluid dynamics with adiabatic index $\gamma=4/3$.   For the progenitor structure we adopt a complete $n=3$ polytrope, so that\footnote{Using the standard definitions for polytropes, $\tilde{\rho}_h = \xi_1^{n+1} |d\theta/d\xi|_1^{n-1}/4\pi$.} $\tilde{\rho}_h=0.3242$ and $\tsehat= 0.451t_*$.  For numerical stability we embed the progenitor in an extended region of very low density.  We ensure that the background density is several orders of magnitude lower than any resolved region of the progenitor, which sets an upper limit of $\sim 10^{-12}\rho_*$ in practice.   Consistent with $\Pi_{\rm grav}\lll1$ we adopt a uniform, negligible pressure in the initial state (also required for numerical stability).  We ensure that the initial sound speed is not too large in the ambient medium, as this leads to inefficiently small time steps.  

Within our progenitor we launch an aspherical explosion in two steps.  First, we add a finite thermal energy into an inner region ($\tilde r<0.05$), and allow the resulting blast wave to expand to a radius $\tilde r_a$.  At this point we set all non-vertical components of the velocity to zero (a slight departure from \citeauthor{Afs18}), and then allow the explosion to evolve as a non-spherical event with definite energy $E$.   Within this procedure, the degree of non-sphericity is controlled by the choice of transition radius; following \citeauthor{Afs18} we choose $\tilde r_a=0.12$, and find that this yields $\epsilon = 0.265$ within the fiducial simulation discussed below.   Evaluating equation (\ref{eq:diffpstar}) we find that the global oblique diffusion parameter within this simulation is  $\diffpstar = 176$ $\Pi_{\rm diff}$. 

We depart from \cite{Afs18}  in the choice of code and numerical algorithm, as we use Lagrangian moving-mesh code HUJI-RICH rather than the Eulerian code Flash.  Our choice comes with several advantages.  We can keep the overall problem size small enough so that simulations can be completed in a few hours on a personal laptop.   We can nevertheless achieve sufficiently high resolution in the sub-surface layers to resolve $\ell_p$ over most of the stellar surface. This  mass resolution advects with the flow, so the ejecta collision is also highly resolved.  Further, by coarsening the initial resolution with distance we are able to define an ambient region of nearly arbitrarily large radius. Finally, using a Lagrangian method allows us to track quantities along the trajectories of individual fluid elements, a feature we will exploit in our analysis.     

To fully benefit from these advantages we require a smoothly varying resolution with the desired properties.  After some experimentation, we adopted an initial radial cell spacing function
\begin{equation} \label{eq:delta-r} 
  \delta \tilde r(\tilde{r})  = \left\{
        \begin{array}{lcl}
            (1-\tilde r) \delta \tilde r_c + \tilde r \,\delta\tilde r_s &~~& \tilde r<1 \\
           \min\left[\delta\tilde r_s + (2 \alpha - \delta\tilde r_s)(\tilde r - 1),\ \alpha \tilde r\right] &~~&  \tilde r \geq 1.
        \end{array}
    \right.
\end{equation}
with fiducial values  $\delta \tilde r_c = 0.025$,   $\delta \tilde r_s = 0.001$,  and $\alpha = 0.025$.  We fix the initial resolution to be $\delta r(r)$ in the polar direction as well, by defining a spiral of mesh points with the appropriate winding angle.  With these choices the initialization radius  $r_a$ is resolved by six cells (approximately $\tilde r_a/\delta\tilde r_c$), while the cells we track in \S~\ref{sec:anl}, which originate from $\tilde{r}=0.97$, are separated from the surface by 23 cells (approximately $0.03/\delta\tilde r_s$) so that their non-radial acceleration will not be affected by finite resolution.  Finally,  $\delta \tilde{r}$ transitions continuously to a moderate relative resolution at radii well outside the star ($\alpha^{-1}=40$ cells per e-folding of $r$), allowing us to capture the equatorial collision within a large volume.  Our fiducial parameters give an outcome consistent with \citet{Afs18}, at  lower computational cost, while also capturing larger collision radii.

\begin{figure*}[ht!]
\gridline{\fig{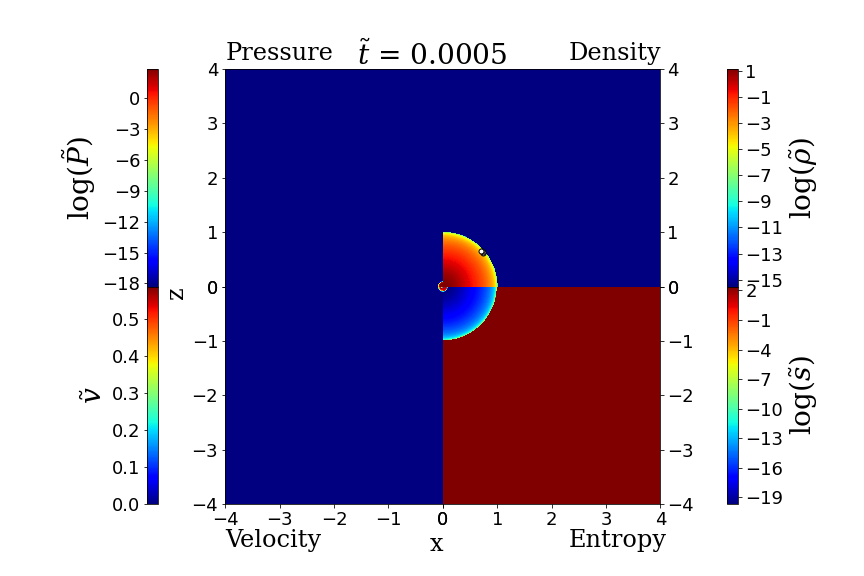}{0.5\textwidth}{(a) Initialization}
          \fig{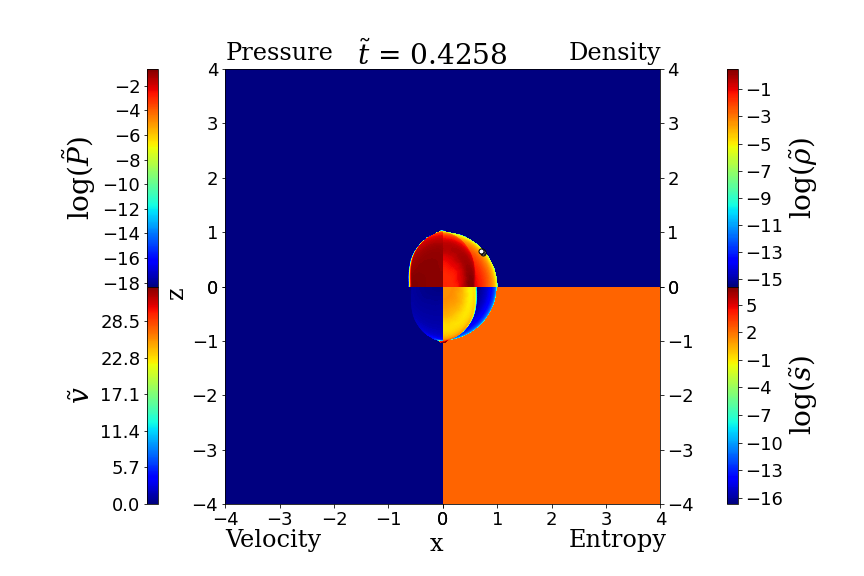}{0.5\textwidth}{(b) Shock breakout at poles}}
\gridline{\fig{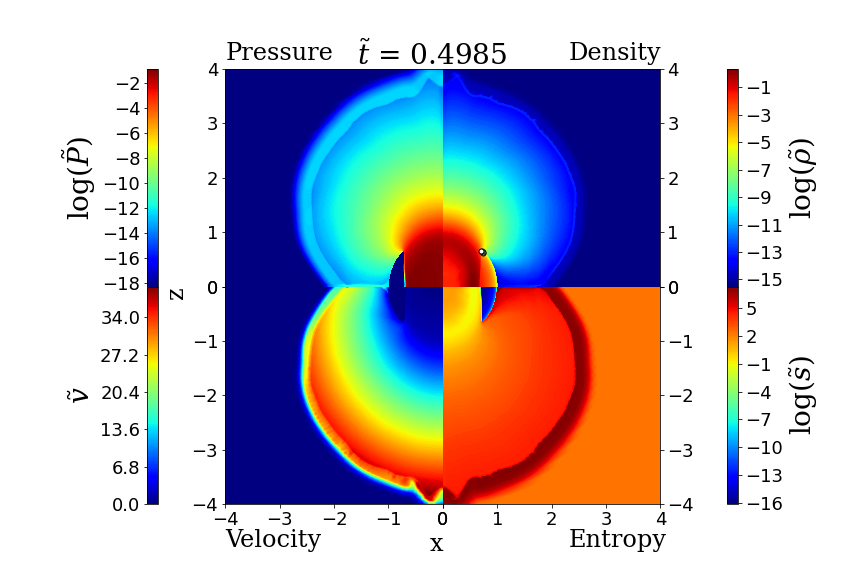}{0.5\textwidth}{(c) Oblique breakout develops}
          \fig{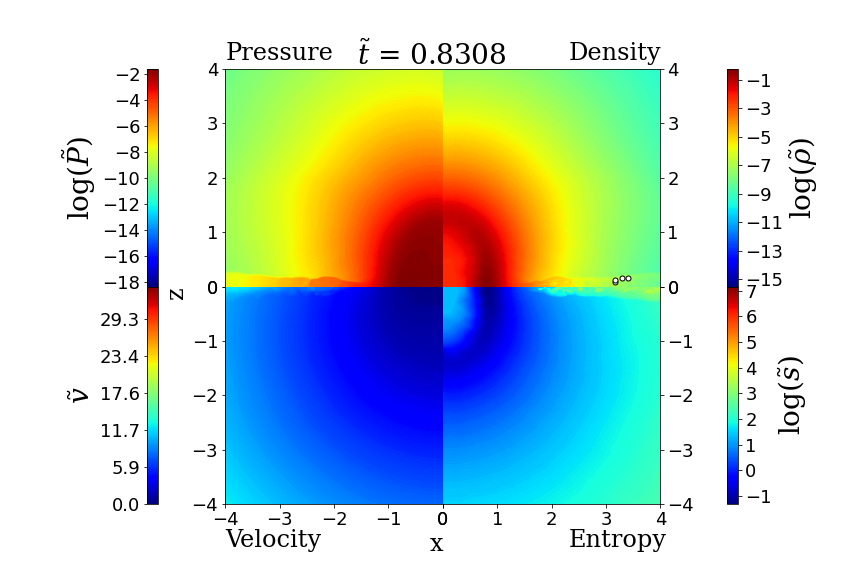}{0.5\textwidth}{(d) Circumstellar collision within ejecta}}
\caption{ Snapshots of pressure, density, entropy parameter $P/\rho^\gamma$, and velocity at four representative stages of the non-spherical evolution.   Ejecta collide in an equatorial region that becomes visible in the last two panels.  White dots represent fluid elements that are tracked to assess their diffusion properties. 
\label{fig:evo}  }  
\end{figure*}

Figure \ref{fig:evo} displays the evolution of fluid quantities at four stages of this fiducial explosion: initialization, incipient breakout, developing oblique shock forward-reverse shock interaction with ambient material, and post-explosion expansion.  The ejecta collision is visible in the equatorial region of the final two panels.  

We note that the feature visible near the axis in the third panel is numerical rather than physical in nature, as it changes with numerical resolution.  We speculate that it reflects  discretization artifacts arising from resolution of the initialization zone $r_a$, of the obliquity scale $\ell_p$ in regions near the axis, or of the low-density ambient material, or  some combination of these effects.  It does not appear to affect our results, which concern flow from mid-latitudes to the equator.

\begin{figure}[ht!]
\epsscale{0.91}
\plotone{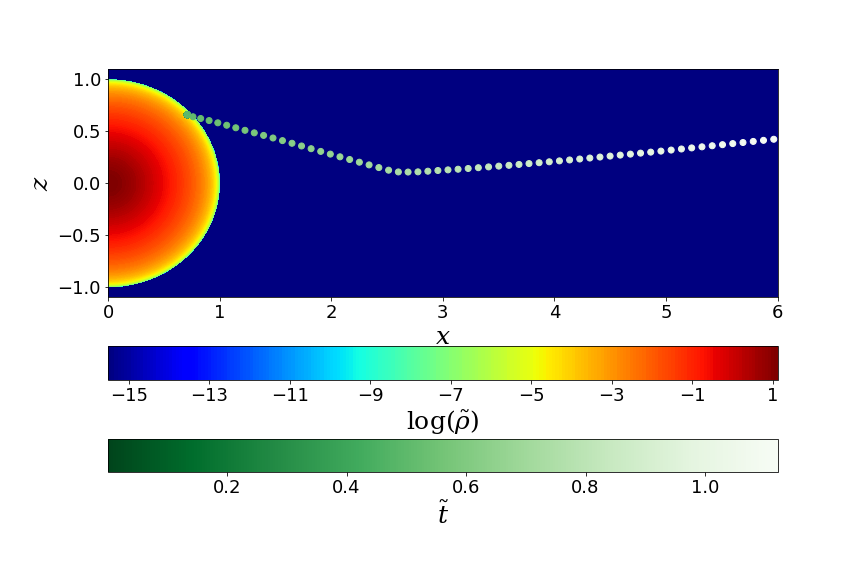}
\caption{The track of a single fluid element, plotted in relation to the initial stellar density profile.
\label{fig:cell_path}}
\end{figure}

Within the simulation we choose a number of zones in the outer stellar envelope ($\tilde{r} = 0.97$, various polar angles $\theta$) for tracking and analysis.  Figure \ref{fig:cell_path} shows the trajectory of  one such cell, showing its acceleration onto a non-radial trajectory, its interaction with the equatorial collision zone, and its redirection onto a new path.  

\section{Results and Analysis} \label{sec:anl}

We seek to compare the strength of radiation diffusion between two phases in the life of a fluid element: its acceleration onto a non-radial trajectory, and its collision with other ejecta.  For this we will need a measure of diffusion strength.  A logical choice, as with $\diffpstar$, is the inverse P\'eclet number, (diffusivity)/[(length)$\cdot$(velocity)], for appropriate choices of length and velocity.   Because the  fluid velocity is not Galiean invariant (and invariance is desirable: see \citealt{Afs18}), we prefer the dimensionally equivalent ratio (diffusivity)$\cdot$(time)/(length)$^2$, for appropriate choices of length and time.   

The appropriate time, for fluid element $j$ that is crossed by the shock at time $t_{s,j}$, is the time since this occurred: $\Delta t_j = t-t_{s,j}$.   For the length, we consider two options.  One we take to be the maximum of the element's initial depth and its current altitude above the stellar surface: we call this the `geometric' length, $L_{{\rm geom}}(t) = \max[R-r_{0,j}, r_j(t)-R]$, and define the corresponding diffusion measure 
\begin{equation} \mathfrak{d}_{{\rm geom},j}(t) = \frac{\nu_j(t) \,\Delta t_j} {L_{{\rm geom},j}(t)^2}
= \frac{\Delta \tilde{t}_j}{3\tilde \rho_j(t) \tilde L_{{\rm geom},j}(t)^2 } \Pi_{\rm diff} .\end{equation} 
Another, purely local estimate of the relevant physical length is the pressure scale height $L_{P}(t) = P/|\nabla P|$, which can be evaluated as $L_P(t) = P/(\rho a)$, using the acceleration ${\mathbf a} = -(\nabla P)/\rho$.   This defines 
\begin{equation} \mathfrak{d}_{P,j}(t)= \frac{ \nu_j(t)\,\Delta t_j} {L_{P,j}(t)^2}
= \frac{\Delta \tilde{t}_j}{3\tilde \rho_j(t) \tilde L_{P,j}(t)^2 } \Pi_{\rm diff} .\end{equation} 

\begin{figure}[ht!]
\plotone{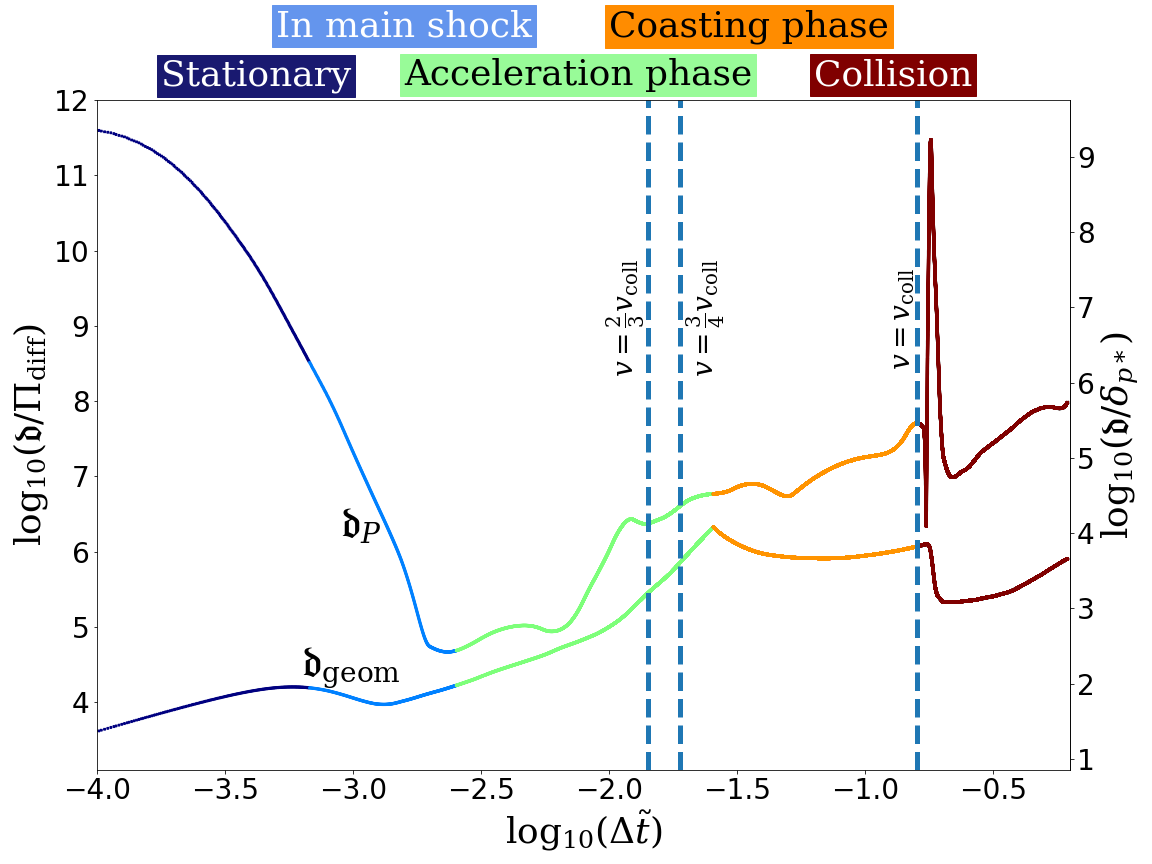}
\caption{Evolution of the local diffusion parameters defined in \S\,\ref{sec:anl}, normalized to the global diffusion measure $\Pi_{\rm diff}$, for a single element originating at $\tilde{r}_0=0.97$ and $\theta_0= 40^{\circ}$ with evolutionary phases labeled.  Vertical dashed lines indicate the times we choose for reference: just before crossing into the shocked collision zone (rightmost dashed line), and partway through the post-shock phase of acceleration, when the velocity is either 2/3 or 3/4 of its pre-collision value.  The change in $\mathfrak{d}_P$ and $\mathfrak{d}_{\rm geom}$ between reference times indicates the potential for a collision to form and also release photons. 
\label{fig:anl}}
\end{figure}

In Figure \ref{fig:anl} we show the evolution of $\mathfrak{d}_{{\rm geom},j}(t)/\Pi_{\rm diff}$ and $\mathfrak{d}_{P,j}(t)/\Pi_{\rm diff}$ for a representative fluid element, the one whose trajectory is displayed in Figure \ref{fig:cell_path}.    

We distinguish five phases of the cell's evolution, as follows.  1:\ Shocking, in which the element is crossed by the shock.  We identify this as the period during which it has moved fewer than three initial resolution elements.  2:\ Post-shock acceleration, during which it gains speed while moving a distance comparable to its initial depth.   
3:\ Coasting, in which the acceleration is relatively low.  4:\ Collision, initiated when the element crosses a shock associated with the collision region.   5:\ Rebound, a second coasting phase with the opposite vertical velocity.   

The crux of our analysis is to evaluate how much  a diffusion parameter ($\mathfrak{d}_{{\rm geom},j}$ or $\mathfrak{d}_{P,j}$) changes between the period of acceleration and when it enters the collision zone (at which point the diffusion parameters reflect the surrounding ejecta, rather than the properties of the collision zone itself).  While the latter reference time is unambiguous, the former is less distinct.  In \autoref{fig:anl} we show two possible choices for the earlier reference time: when the fluid element's velocity is either $\frac23$ or $\frac34$ of its velocity at the collision shock.  (With either choice, its heading is consistent to within 4$^\circ$ over the entire reference period.)   

As is clear from the figure, for the element plotted,  $\mathfrak{d}_{P,j}$ increases by just over an order of magnitude between the two reference times,  while $\mathfrak{d}_{{\rm geom},j}$ increases much less.  
Specifically, the change in $\log_{10} \mathfrak{d}_{P,j}$ for this fluid element is (1.34, 1.10) when the initial reference velocity is (2/3, 3/4) of the final value, while the change in $\log_{10}\mathfrak{d}_{\rm Geom,j}$ is (0.62, 0.22).  

While the ambiguity of initial reference time and the choice of diffusion measure introduces some uncertainty into our conclusions, taking the geometric mean of all four estimates suggests an increase of $\sim$0.82\,dex (a factor of $\sim$6.6), in diffusivity between the acceleration and collision regimes.

\section{Discussion} \label{sec:conc}

We have found that the fluid elements which can participate in circumstellar collisions do, in fact, experience an increase in the dimensionless diffusion parameter (inverse P\'eclet number) between the acceleration and collision phases -- although the degree to which this is true depends on which length scale we use in the evaluation, and also on the criterion we use to identify when a fluid element is sufficiently deflected to undergo a collision.  Our estimates for the change in local diffusion parameter vary from less than a factor of two to over an order of magnitude, with a best-guess estimate of 0.82 dex, a factor of 6.6. This indicates the strong possibility that real explosions can both successfully create colliding ejecta flows, and release the light created in the collisions, if they inhabit the narrow range of the global parameter $\diffpstar$ for which diffusion is weak at first, and then becomes strong. 

Based on the range of local diffusion parameters we observe for fluid elements like the one tracked in Figure \ref{fig:anl}, and based on the ratio $\diffpstar/\Pi_{\rm diff}$ characterizing this explosion, we estimate that the most conducive conditions are in the vicinity of $\diffpstar = 10^{-4.5}$.   The optimal value of $\diffpstar$ depends on the fluid element under consideration, however, so there is likely to be a portion of the flow that engages in this behavior over a wider range of parameters.   We encourage  exploration using radiation hydrodynamics simulations to test whether the corresponding explosions can indeed produce circumstellar flashes.  To be useful, such simulations will need to resolve the obliquity depth $\ell_{p*}$, and also encompass multiple stellar radii to include the collision zone.

To translate our result into practical terms, let us identify the range of explosion parameters for which the fluid element tracked in Figure \ref{fig:anl} is likely to collide and release photons.    Given the sensitivity to eccentricity evident in equation (\ref{eq:diffpstar}), an order of magnitude in $\diffpstar$ translates into  a factor of only $\sim 1.45$ over which $\epsilon$ can vary, within a given explosion (all else fixed).    For reference, using the $n=3$ polytropic atmosphere,  $\diffpstar=10^{-5}\rightarrow 10^{-4}$ when 
\begin{equation}
\label{eq:ep_range}
\epsilon =  \left(0.16\rightarrow 0.23\right) \left[\frac{ R_{10}^2 (1-\tilde\ell_{p*})^{3}  \tilde\rho_h^{1/3}}{
E_{51}^{1/2} M_{10}^{1/2}  \kappa_{0.2} }\right]^{0.162}, 
\end{equation} 
where  parameters are normalized according to their subscripts ($10^{51}$\,erg for $E$, $10\,M_\odot$ for $M$, $10\,R_\odot$ for $R$, and $0.2\,$cm$^2$\,g$^{-1}$ for $\kappa$).  This  range of $\epsilon$ is plausibly achieved by SESNe, especially considering their compact sizes and conspicuous signatures of non-sphericity; and the relative insensitivity to $R$ suggests that a subset of Type IIb explosions may generate excess emission if they can be sufficiently aspherical.  

When considering the detectability of collision-induced x-ray and UV flashes by future observatories such as Einstein Probe \citep{zhang2022first} or ULTRASAT \citep{2022SPIE12181E..05B}, it is worth remembering that the fraction of $E$ directed into the vertical kinetic energy that can drive such an event is itself a strong increasing function of $\epsilon$; \cite{S14} estimate this fraction to be $\sim$0.33$\epsilon^{4.8}$ for an $n=3$ polytropic model.   This  limits the expected rate of detections and favors the largest, most aspherical explosions capable of producing a collision flash.

\subsection{Application to SN 2008D} 
The type Ib supernova 2008D was heralded by an unusual, several-hundred-second long x-ray flash (XT 080109) with a radiated energy of $\sim 2\times10^{46}$\,erg \citep{Soderberg08_SN08D}. It is worthwhile to examine whether a circumstellar collision transient could have produced this event, as suggested by \citet{cou11} and \citet{M13}.  Because the energy-to-mass ratio is well constrained by the photospheric velocity at maximum light via $\sqrt{2} v_*\simeq v_{\rm max}$ \citep{dessart16vphot}, we choose $v_*=8,000 f_v$\,km\,s$^{-1}$ with $f_v\simeq1$ \citep{Soderberg08_SN08D}; the ejected mass is then $3.1  M_\odot$ per $4.0\times10^{51}f_v^2$\, erg of kinetic energy.  Because the x-ray flash rises to a peak in 100\,s and that 90\% of its fluence emerges in $\sim 300$\,s \citep{Soderberg08_SN08D}, we adopt 300$f_t$\,s for its duration, and associate this with the period of high kinetic luminosity in the circumstellar collision. We note that \cite{Modjaz2009} infer a lower energy ($0.6\times 10^{46}$ erg) with their spectral fit, as well as a longer duration ($f_t =1.57$).  In the fiducial simulation of \citet{Afs18}, for which $\epsilon = 0.26$, the collision lasts from roughly $0.4t_*$ to $1.4t_*$, and we expect its duration to scale in proportion to $\epsilon$.  Combining these, $300 f_t\,s = 1.0 (\epsilon/0.26) t_*$.  The stellar radius is then $R=3.4 f_t f_v (\epsilon/0.26) R_\odot$.  As previously noted by \citet{cou11}, the progenitor can be significantly more compact in an aspherical model than in a spherical one. 

Our new constraint on $\epsilon$ arises from the requirement that the collision must both exist and radiate promptly.  Letting $ f_\epsilon $ measure $\epsilon$ relative to the geometric mean of the range in equation (\ref{eq:ep_range}), our constraint is  $0.83<f_\epsilon<1.2$. 

Can an explosion that satisfies this condition also produce the observed fluence?  Estimating the available energy as $0.33\epsilon^{4.8}E$ \citep{S14}, and taking the weak parameters $\tilde{\rho}_h$,  $(1-\tilde\ell_p)^3$, and $\kappa/0.2$\,cm$^2$g$^{-1}$ all to be unity, we find the kinetic energy crossing the collision zone to be
\begin{equation}  \label{eq:EcollSN08D}
 E_{\rm coll} = 2.2 \times 10^{46} \left(E_{51}/3\right)^{-0.16} f_\epsilon^{7.1}  f_v^{3.5} f_t^{2.3} \,{\rm erg}. 
\end{equation}
We note that the term $f_\epsilon^{7.1}$ varies from $1/3.6$ to $3.6$ across the range quoted in equation (\ref{eq:ep_range}).  The very weak dependence on $E_{51}$ is particularly intriguing.   However, the combination $f_v^{3.1} f_t^{2.3}$ introduces a strong dependence on these somewhat uncertain parameters. 

The result in equation (\ref{eq:EcollSN08D}) roughly matches the radiated energy of XT\,080109, which suggests that SN\,2008D's prompt x-ray flare could well have been the result of an equatorial collision among non-radial ejecta, assuming efficient conversion into 2-10 keV photons. This requires $\epsilon \simeq 0.11$, which is plausible given that the optical spectra of SN 2008D show evidence for asphericity in both the high- and low-velocity ejecta \citep{Modjaz2009}. Further modeling will be necessary to evaluate this scenario in comparison to spherical models that invoke a thick circumstellar environment \citep[e.g.][]{Soderberg08_SN08D, chevalierfransson08,SN2008Svirski}.

\begin{acknowledgments}
We thank Stephen Ro, Amalia Karalis, and Shannon Bowes for helpful suggestions and comments.
We are especially grateful to the referee for suggesting a detailed examination of SN\,2008D. 
B.S.'s and C.D.M.'s research are supported by an Natural Sciences and Engineering
Research Coucil of Canada (NSERC) Discovery Grant, and B.S.  was additionally supported by an NSERC Summer Undergraduate Scholarship and by the hospitality and mentorship of the Summer Undergraduate Research Program at the Dunlap Institute in Toronto.   A.Y. was supported by a Canadian Institute for Theoretical Astrophysics (CITA) Postdoctoral Fellowship.  
\end{acknowledgments}

 \noindent {\em Data Availability: }All data in this paper are available upon reasonable request to the corresponding author.

\bibliography{sources}{}
\bibliographystyle{aasjournal}

\end{document}